# Design of nematic liquid crystals to control microscale dynamics

*Oleg D. Lavrentovich*

*Advanced Materials and Liquid Crystal Institute, Department of Physics, Chemical Physics Interdisciplinary Program, Kent State University, Kent, OH 444240, USA; ORCID 0000-0002-0128-0708*

**Keywords:** particle-like solitary waves in nematics; liquid crystal enabled electrokinetics, patterned liquid crystal elastomers, living liquid crystals

## Summary

Dynamics of small particles, both living such as swimming bacteria and inanimate, such as colloidal spheres, has fascinated scientists for centuries. If one could learn how to control and streamline their chaotic motion, that would open technological opportunities in areas such as the transformation of stored or environmental energy into systematic motion, micro-robotics, and transport of matter at the microscale. This overview presents an approach to command microscale dynamics by replacing an isotropic medium such as water with an anisotropic fluid, a nematic liquid crystal. Orientational order leads to new dynamic effects, such as propagation of particle-like solitary waves. Many of these effects are still awaiting their detailed mathematical description. By using plasmonic metamask photoalignment, the nematic director can be patterned into predesigned structures that control dynamics of inanimate particles through the liquid crystal enabled nonlinear electrokinetics. Moreover, plasmonic patterning of liquid crystals allows one to command the dynamics of swimming bacteria, guiding their trajectories, polarity of swimming, and concentration in space. The patterned director design can also be extended to liquid crystal elastomers, in which case the director gradients define the dynamic profile of elastomer coatings. Some of these systems form an experimental playground for the exploration of out-of-equilibrium active matter, in which the levels of activity, degree of orientational order and patterns of alignment can all be controlled independently of each other.

*Author for correspondence (xxxx@yyyz.zz.zz ).
†Present address
Department, Institution, Address, City, Code, Country

# Main Text

**INTRODUCTION**

Scientific exploration of soft matter has provided the basis of truly transformative and life-changing technologies, such as liquid crystal displays (LCDs), development of new approaches to drug delivery, sensors, nano-templated materials, etc. A unifying theme of past soft matter research and development has been the achievement of monocrystal behavior, as exemplified by a uniform alignment of molecules in the pixels of an LCD screen and their collective realignment by an externally applied electric field. The new frontier of soft matter research, the one that will define the long-term development of fundamental science and technologies, is to learn how to translate nanoscale molecular organization into the mesoscale soft matter with spatially-non-uniform yet organized, functional and active structures, with tunable local and non-local responses to external cues and dynamic collective behavior coordinated in space and time. Active systems comprised of interacting units that use internal sources of energy to propel themselves are of especial interest and challenge since they resemble living systems.

Mathematical modelling of this new frontier of soft matter science is expected to focus on understanding how spatially-varying molecular patterns can be used to design, produce, and command the structure, properties, and dynamics of matter with a hierarchy of length scales, from a few nanometers, the typical size of an organic molecule, to the scale at which the structure reveals itself to the macroscopic world, for example, through interactions with biological organisms or synthetic microbots. Recent advances in understanding active matter (1-4) provide a compelling evidence of the importance of widening the scope of mathematical modelling to include out-of-equilibrium behaviors, driven either by internally stored sources of energy or by harvesting the energy of the environment.

One of the interesting directions is to explore dynamics at the microscale, which usually corresponds to the small Reynolds number regime, i.e., to the prevalence of viscosity over inertia. Dynamics of small particles in fluids has fascinated scientists for centuries, since van Leeuwenhoek observed in 1674 tiny creatures, nowadays known as "bacteria", swimming chaotically in a droplet of water. Much later, Brown found that even inanimate small particles, when placed in water, engage in a somewhat similar chaotic dynamics. Control of the dynamic behavior of small particles could enable new interesting applications in the energy transformation, micro-robotics, transport of matter, etc. Usually, the microscale dynamics is explored for an isotropic environment such as water. In an 3isotropic medium, the dynamics can be controlled either by externally applied electric fields or by gradients of some properties, such as the concentration of nutrients in the case of bacteria. The two approaches are certainly valuable, but they are often limited, for example, by a temporary character of the gradients. It is thus of interest to explore whether the microscale dynamics can be controlled by replacing an isotropic environment with a liquid crystalline fluid, thus providing some sense of direction associated with the orientational order of the material. This brief review focuses on the recent advances in designing the simplest type of the liquid crystals, the so-called nematic, as a medium to guide microscale dynamics. It is based on the lectures presented by the author at the Spring 2019 School on The Mathematical Design of Materials at The Isaac Newton Institute. The presented examples deal with particle-like solitons of the director field (Section I), liquid crystal-enabled electrokinetics (Section II), liquid crystal elastomers (Section III), and living liquid crystals, representing microswimmers such as bacteria dispersed a in non-toxic water-based lyotropic nematic (Section IV). Most of the themes emerged only recently through experimental studies and thus still await a thorough mathematical description.

## I. ELECTRICALY POWERED PARTICLE-LIKE SOLITONS IN NEMATICS

The theme of electric field effects in nematics appears to be rather well understood since its most celebrated realization, the so-called Frederiks effect, is widely used in modern informational liquid crystal displays (5, 6). The field-induced reorientation of the nematic molecules described by the director $\hat{\mathbf{n}}$, is either homogeneous over the area of electrodes, as in displays (5), or periodically modulated, as in the phenomenon called electroconvection (5, 7) similar to Rayleigh-Bénard thermal convection (8). Theoretical description of the director response to an electric field is difficult because of numerous mechanisms involved, such as anisotropic bulk elasticity and surface "anchoring" of the director, anisotropy of dielectric permittivity, electric conductivity and viscous drag, surface and flexoelectric polarization, spatial charge build-up at director gradients, injection of current carriers from electrodes, etc. The pioneering works by V. Frederiks, P.G. de Gennes, R.B. Meyer, W. Helfrich and others helped to develop current understanding of the electric field effects in liquid crystals. However, because of the above-mentioned difficulties, the theoretical models are usually reduced to one-dimensional or two-dimensional settings, often with an assumed periodicity, as, for example, in the treatment of one- and two-dimensionally periodic electrohydrodynamic instabilities (5-7, 9, 10). An intriguing question is whether the electric field can create spatially localized perturbations, such as solitons of molecular reorientation. Since the word "soliton" is used in many different contexts, below

*Phil. Trans. R. Soc. A.*

I attempt to separate different varieties of soliton objects met in liquid crystals. The most important distinction is of a topological nature, as the solitons can be topologically protected (non-transformable into a ground uniform state by continuous deformation of the director) or unprotected, topologically equivalent to the uniform ground state. I first describe the current status of the so-called topological solitons and nematicons and then discuss the topologically trivial dissipative solitons, represented by the so-called director bullets (11-14) or directrons (14).

**Topological solitons**. Early studies of spatially restricted director perturbations in nematics, termed topological solitons or topological configurations (15), started about 50 years ago with the discussion of *static* linear and planar solitons produced by the magnetic or electric fields (16). These solitons form when the external field aligns the director parallel to itself because of dielectric or diamagnetic anisotropy. Since the nematic is nonpolar, the states $\hat{\mathbf{n}}$ and $-\hat{\mathbf{n}}$ that are both parallel to the applied field, are equivalent to each other. If both states are trapped by the applied field, there must be a transition region in which $\hat{\mathbf{n}}$ reorients by $\pi$. This region of a smooth director reorientation of a finite width (determined by a balance of elasticity and coupling to the field), represents a static linear or planar topological soliton (15). Multidimensional static topological solitons, such as three-dimensional (3D) particle-like nonsingular perturbations of the director field $\hat{\mathbf{n}}(\mathbf{r})$ are deemed unstable with respect to shrinking. The reason is easy to understand by considering a hypothetical static particle soliton in a nematic. A decrease in size, $L \to \lambda L$ by a factor $\lambda < 1$ entails an increase of the elastic energy of director distortions $\propto 1/L^2$ by a factor $1/\lambda^2$, and a decrease in the topological soliton's volume by a factor of $\lambda^3$, thus the total elastic energy decreases as the soliton shrinks, $F \to \lambda F$ (15). The instability of multidimensional solitons in nonlinear field models is generally known as Derrick-Hobart theorem (17). A particular mechanism of stabilization of static 3D topological solitons in liquid crystals is provided by helical twisting of the director, when the nematic is doped with chiral molecules and becomes a cholesteric. A similar Dzyaloshinskii-Moriya mechanism is known in the theory of helicoidal structures in ferromagnets (18). Stable particle-like topological structures in cholesterics were observed by Haas and Adams (19, 20), Kawachi et al (21), Bouligand et al (22), Smalyukh et al (23-27), Guo et al (28), Nych et al (29), Posnjak et al (30). These formations are static and do not require any motion (although they can be forced to move by an external field, as demonstrated by the Smalyukh's group (25, 26)). The stability of static topological solitons in cholesterics against shrinking is guaranteed by the helicoidal structure that maintains a fixed pitch. Regions of $\hat{\mathbf{n}}$ and $-\hat{\mathbf{n}}$ that differ by a rotation of the director by $\pi$ are separated by half of the pitch that cannot be moved closer to each other without a dramatic increase of the elastic energy of the director twist (22, 31, 32). Cholesterics show an extraordinary broad range of field-, surface anchoring-, and light-generated topologically nontrivial solitons, including structures that resemble skyrmions. The term "skyrmion" embraces a broad range of topological solitons in various fields of physics, including ferromagnets and cholesterics (24, 25, 28, 33, 34) and first considered by Skyrme (35) as a model of elementary particles (33). For a detailed description of skyrmions and other topological solitons, see the textbook by Manton and Sutcliffe (33). Their mathematical description is usually performed in the so-called one-constant approximation (32, 36-42) in which all elastic constants describing various deformations (splay, twist, bend, saddle-splay) of a vector field (such as the director) are assumed to be equal to each other; lifting this assumption might be an interesting task for mathematical modelling, potentially leading to a discovery of new structures and symmetries.

**Nematicons**. Current studies on light-induced solitons in nematics mostly deal with optical solitons (43-46) representing propagating self-focused laser beams, called "nematicons" (43, 46). Studies of nematicons are a part of very broad research on optical solitons in different nonlinear media (47-52), see reviews by Malomed et al (53, 54). Optical solitons are usually described using a nomenclature $(m+1)\mathrm{D}$, which means that the light beam can diffract in $m$ dimensions as it propagates in one dimension (47, 53). Of especial interest are the so-called "light bullets" (55) or (3+1)D spatiotemporal solitons, which are self-confined in the longitudinal and both transverse directions and can be used in fast optic-logic systems (56). Multidimensional solitons, unlike their 1D counterparts, are vulnerable to various instabilities and are extremely hard to realize experimentally, see the discussion in Refs (52-54). For example, there are no reports on experimental observations of stable light bullets that can survive collisions without losing energy. Since the director deformations within nematicons are topologically equivalent to the uniform state, these are topologically unprotected formations.

**Dissipative solitons**. Among the broad family of solitons, there is a class of localized externally driven structures, often called dissipative solitons (57-59). Dissipative solitons represent a portion of a pattern surrounded by a homogeneous steady state; below a certain strength of the driver, they vanish (57-59). Experimentally, dissipative solitons were realized in the form of electric current filaments in a 2D planar gas-discharge system (60). As often happens, nematics provide a fertile ground to realize entities that are difficult to form or to observe in other materials: very recently, various types of dissipative solitons driven by an electric field were experimentally realized in nematics (11-14, 61).

Dissipative solitons in electrically driven nematics are in the form of (3+1)D (11) or (3+2)D (12) particle-like solitary waves of the director $\hat{\mathbf{n}}$, called by Li et al (11, 12) "director bullets" in analogy with the "light bullets." A more descriptive term is a "directron", introduced later by Li et al in Ref. (14), as it stresses the dual wave-particle character of the formation, in which the solitary wave of the director behaves as a particle in events such as collisions with other directrons. The first mathematical model of these structures has been proposed recently by Earls and Calderer (62) in the limit of linear approximation, which is a very promising step towards a complete nonlinear description. Pikin pointed a possible role of field-induced injection of electric carriers in the formation of the director bullets (63). Below I list the underlying experimental features that might be helpful for the development of a complete mathematical model.

The nematic in which the directrons were first observed by Li et al (11, 12), 4'-butyl-4-heptyl-bicyclohexyl-4-carbonitrile (CCN-47), Fig.1, is of the (-,-) type, which means that both dielectric and conductivity anisotropies are negative, $\Delta\varepsilon = \varepsilon_\parallel - \varepsilon_\perp < 0$ and $\Delta\sigma = \sigma_\parallel - \sigma_\perp < 0$, respectively; here the subscripts refer to the direction with respect to the director.

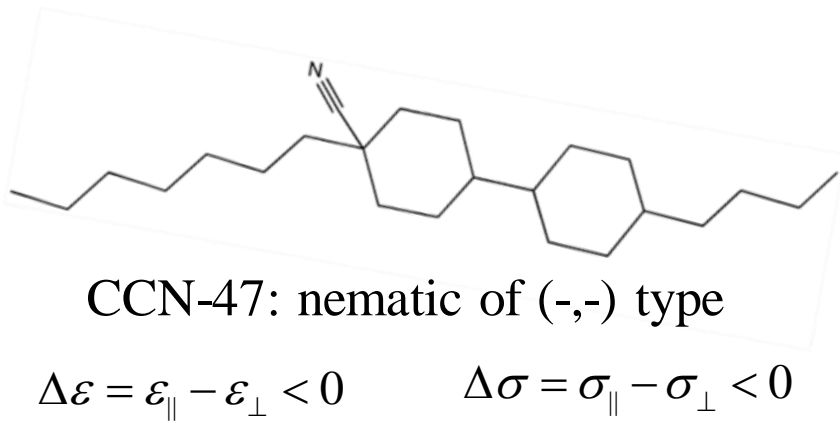

CCN-47: nematic of (-,-) type

$$\Delta\varepsilon = \varepsilon_\parallel - \varepsilon_\perp < 0 \qquad \Delta\sigma = \sigma_\parallel - \sigma_\perp < 0$$

**Fig.1**. Chemical structure of the (-,-) nematic CCN-47 in which directrons are excited by an applied AC electric field.

The experimental conditions under which the directrons are observed are very simple. The cell is prepared in a "planar" fashion, with the director aligned along a single direction in the plane of the cell, $\hat{\mathbf{n}}_0 = (0,1,0)$. A sinusoidal AC electric field $\mathbf{E} = (0,0,E)$ is applied across a cell, bounded by two glass plates with transparent electrodes at the inner surfaces. Once the AC electric field of a certain frequency and an amplitude above some threshold is applied, the system develops spatially-confined perturbations of the director field $\hat{\mathbf{n}}(\mathbf{r})$ that coexist with a uniform director state $\hat{\mathbf{n}}_0 = (0,1,0) = \text{const}$, Figs.2-5. These perturbations are localized along all three spatial dimensions and do not spread while moving over macroscopic distances (11). In the $xy$ plane of propagation, the width and length of the soliton are thousands times smaller than the corresponding dimension of the system. Along the $z$-axis, the stability of solitons is assisted by the surface anchoring at the bounding plates. Within a directron, the director perturbation oscillates with the frequency of the applied AC electric field and breaks the fore-aft or left-right symmetry, which results in the directron's propagation, Figs.2-5.

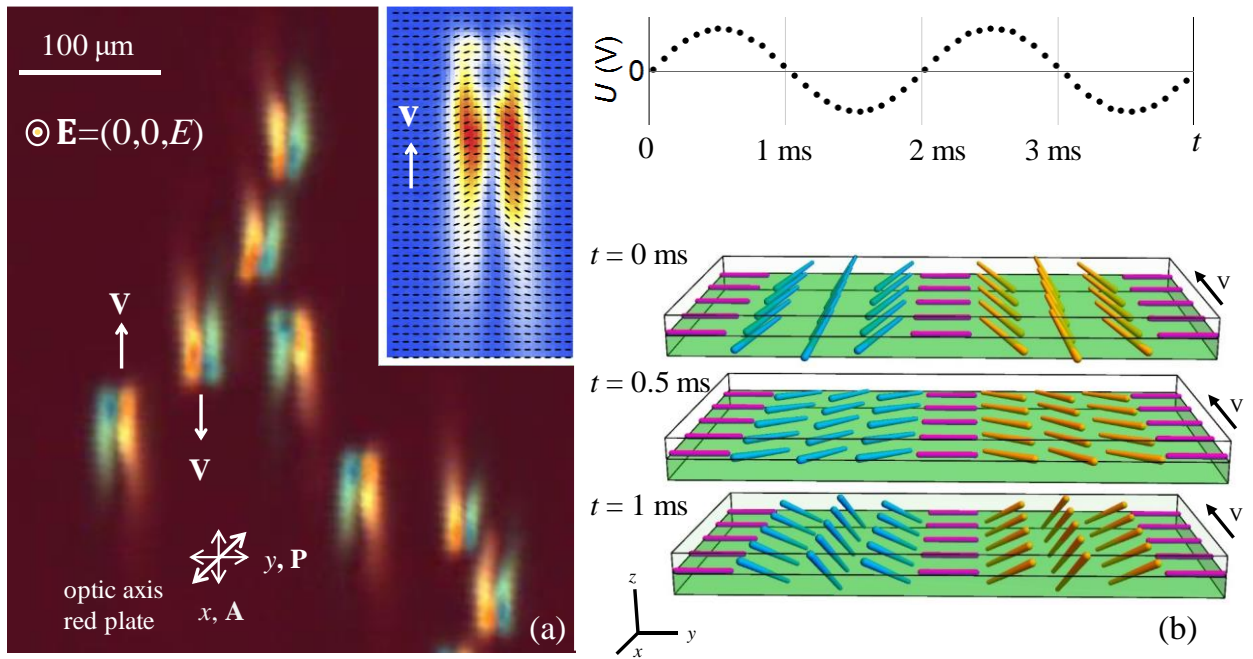

**Fig.2.** Directrons $B^h$ formed in an AC field of a high frequency: (a) optical microscopy texture with the bent director field shown in the inset; (b) director field within the soliton body oscillating up and down with the frequency of the applied voltage U. Redrawn from Li et al (11).

There are different symmetry breaking scenarios that determine the propagation direction and other properties of the directrons. These scenarios depend on the frequency and amplitude of the driving voltage. Directrons induced by fields of high frequency, typically $f > 100$ Hz in CCN-47, abbreviated as $B^h$ directrons, show a left-right asymmetric "bullet" structure when viewed under a polarizing microscope with crossed polarizers, Fig.2a (11). They are formed by an oscillating director deformation predominantly of a bend type, Fig.2b; the bend vector $\mathbf{b} = \hat{\mathbf{n}} \times \text{curl}\,\hat{\mathbf{n}}$ shows a unique direction along the entire body of the $B^h$ directron, Fig.2a and 4a (11).

*Phil. Trans. R. Soc. A.*

The low-frequency ($f <100$ Hz) $B^l$ directrons are of butterfly appearance when viewed in a polarizing microscope, Fig.3a,d, Ref. (12), and exhibit deformation of bend and splay, in which the vectors of bend $\mathbf{b} = \hat{\mathbf{n}} \times \text{curl} \hat{\mathbf{n}}$ and splay $\mathbf{s} = \hat{\mathbf{n}} \text{div} \hat{\mathbf{n}}$ change polarity as one moves along the direction of propagation, Figs.3b,c,e,f and 4b,c. This partial "self-compensation" results in the propagation speed of $B^l$ solitons that is noticeably slower than the speed of high-frequency $B^h$ directrons; in both cases, the velocity increases with the applied field, typically as $E^2$ (11, 12).

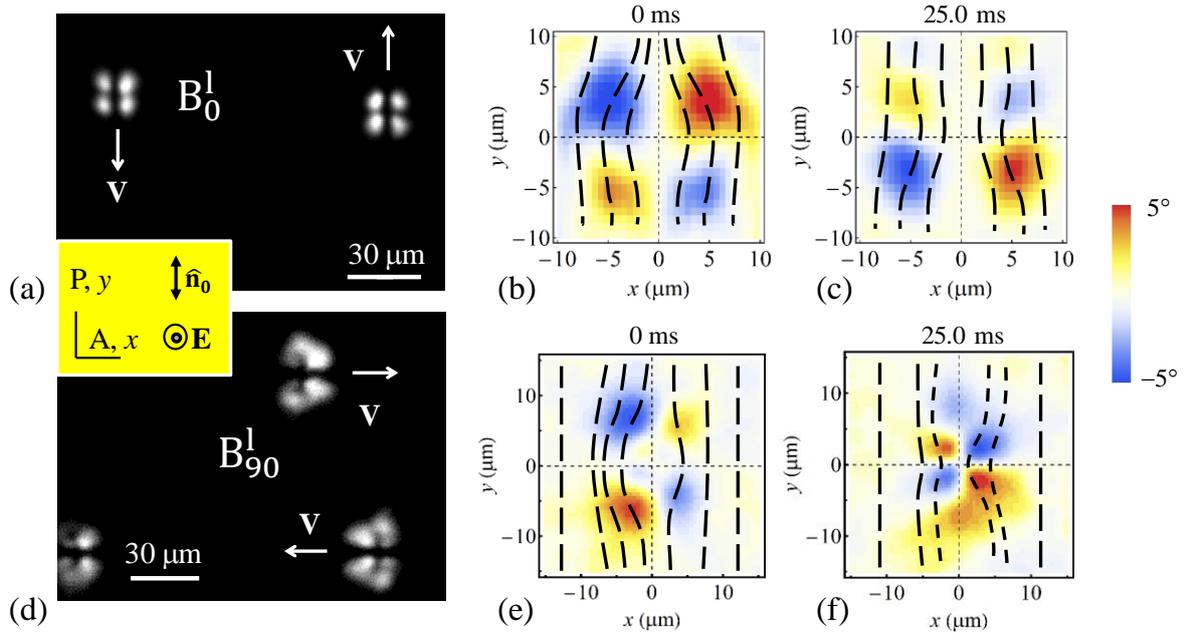

**Fig.3**. Low-frequency directrons $B^l$ that propagate (a,b,c) parallel to the overall director field and (d,e,f) perpendicular to overall director. Instantaneous in-plane director tilt for $B_0^l$ directron (c) at time "0 ms" which corresponds to the minimum of the applied voltage and (d) at time "25 ms" which corresponds to the maximum of a sinusoidal AC field; (e,f) similar director patterns for $B_{90}^l$ directrons. The in-plane tilts are color-coded in accordance with the scale on the right-hand side. The tilts of the dashed lines with respect to the overall director is exaggerated by a factor of 6 for a better clarity. Redrawn from Li et al (12).

The $B^l$ director field, both instantaneous, Fig.3b,c,e,f, and averaged over the period of driving AC electric field, Fig.4b,c, can be conveniently described by introducing a quadrant system, Fig. 4c; the director $\hat{\mathbf{n}}_0 = (0,1,0)$ separates the quadrants I and II. The deformed propagating structure of $B^l$ can show a vertical (containing the *z*-axis) mirror symmetry plane, which passes either between quadrants I and VI, Fig.4b, or between I and II, Fig.4c; there is no mirror symmetry with respect the orthogonal vertical plane (as needed for propulsion). As a result, the corresponding directron propagates along the direction that is either perpendicular, Fig.3b, or parallel, Fig.3c, to the background director $\hat{\mathbf{n}}_0$. If the mirror plane of symmetry is absent, the soliton propagates under some angle to the director $\hat{\mathbf{n}}_0$. An important and interesting property of these $B^l$ directrons with multiple scenarios of symmetry breaking is that a change of the electric field's amplitude and/or frequency steers them in the *xy* plane of the cell, perpendicular to the applied field (12). Since the $B^l$ directrons are limited in space along all three directions and can move along different trajectories within a plane, they are classified as (3+2)D solitons (12).

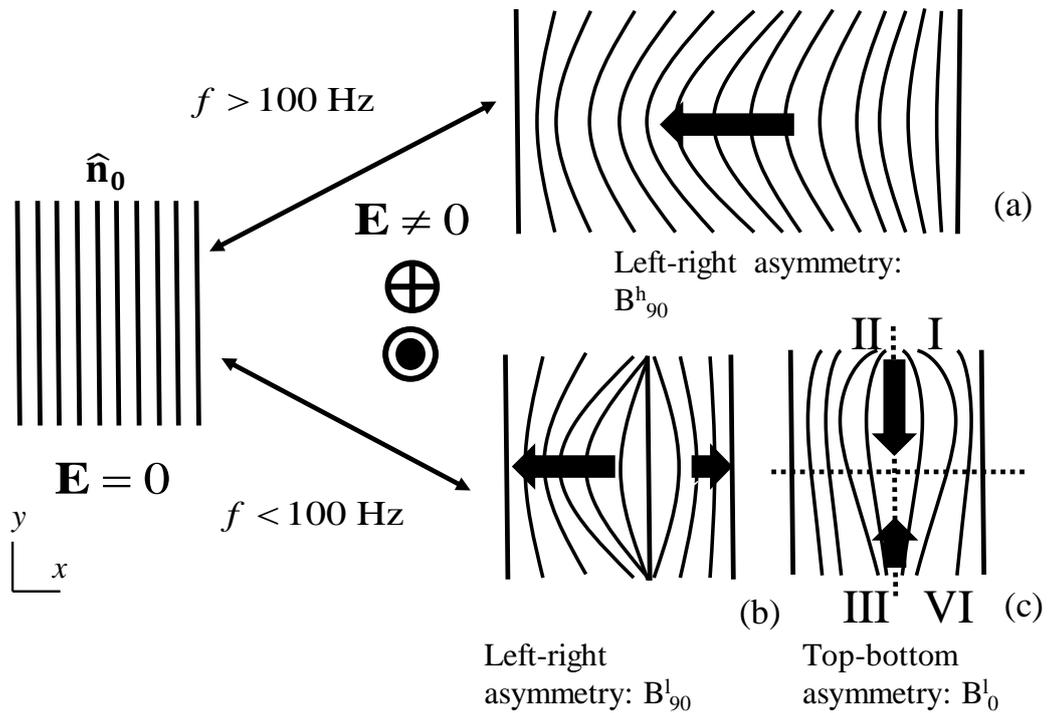

**Fig.4**. Time-averaged asymmetry patterns of directrons driven at high (top row) and low (bottom row) frequencies of the electric field; filled horizontal arrows show bend vector $\mathbf{b}=\hat{\mathbf{n}}\times\text{curl}\hat{\mathbf{n}}$, the vertical arrows show splay vector $\mathbf{s}=\hat{\mathbf{n}}\,\text{div}\hat{\mathbf{n}}$.

The directrons show their particle-like character in collisions, Fig.5. High-frequency $B^h$ directrons typically move perpendicularly to the overall director. When two $B^h$ directrons collide, they briefly unite and then restore their shape and propagation speed, Fig.5 (11). Interestingly, directrons show repulsive/attractive interactions when they move in close proximity. For example, in Fig.5 the post-collision separation of the two colliding $B^h$ directrons along the *x*-axis is larger than before the encounter, which demonstrates repulsive interaction. Besides the recovery scenario illustrated in Fig.5, collisions of two $B^h$ solitons might lead to a number of other outcomes, such as pair annihilation or disappearance of one soliton (11).

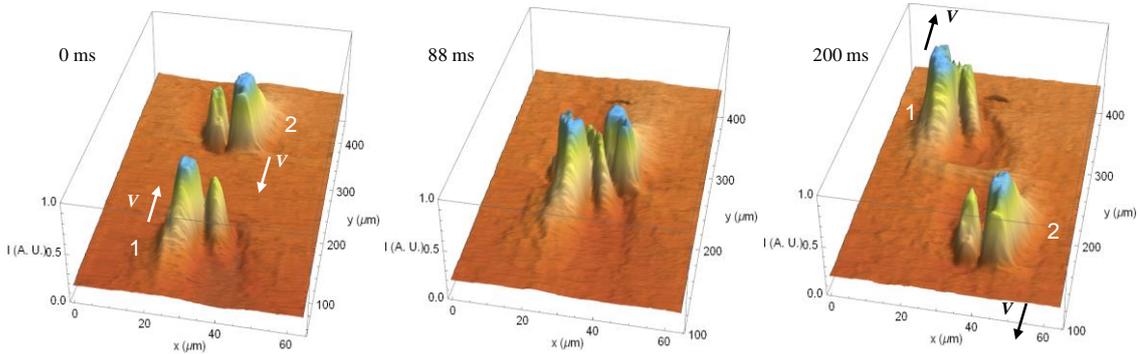

**Fig.5**. Two $B^h$ directrons approaching each other at 0 ms, colliding at 88 ms and moving away after a collision preserving their shapes, at 200 nm. The vertical axis shows the intensity of the transmitted light. Redrawn from Ref. (11).

The "butterfly" $B^l$ low-frequency directrons show even richer behavior in collisions since these solitons can move along different directions (12). For example, in addition to recovery and annihilation, two $B^l$'s can change the direction of propagation by 90° after a collision, or merge into one directron that would move in the direction different from the directions of the original solitons, etc., see Ref. (12) for more details.

The critical frequency that separates the low-frequency $B^l$ directrons from the high-frequency $B^h$ directrons can be associated with the frequency $f_c = \sqrt{\xi^2 - 1}/\tau_M$ introduced by Dubois-Violette et al (64, 65) to describe the so-called conducting and dielectric regimes of electrohydrodynamics in nematics. Here $\tau_M = \varepsilon_0 \varepsilon_\perp / \sigma_\perp$ is the Maxwell relaxation time for a planar cell, $\varepsilon_0 = 8.85 \times 10^{-12}$ F/m, and $\xi^2 = \left(1 - \frac{\sigma_\perp}{\sigma_\parallel} \frac{\varepsilon_\parallel}{\varepsilon_\perp}\right)\left(1 + \frac{\alpha_2}{\eta_c} \frac{\varepsilon_\parallel}{\Delta \varepsilon}\right)$ is the material parameter that depends on conductivities $\sigma_\parallel$ and $\sigma_\perp$, permittivities $\varepsilon_\parallel$ and $\varepsilon_\perp$, viscous coefficients $\alpha_2$ and $\eta_c$. For typical material parameters (11, 12) $\varepsilon_\parallel \approx 5$, $\varepsilon_\perp \approx 8$, $\sigma_\parallel, \sigma_\perp \sim (0.5-2)10^{-8}$ $\Omega^{-1}$m$^{-1}$, assuming $\alpha_2 \approx -\eta_c$, one estimates $\sqrt{\xi^2 - 1} \sim 1$ and $f_c \sim 100$ Hz. In the theory of electrohydrodynamic instabilities, the region $f < f_c$ is associated with the oscillating electric charges (conductive regime), while the region $f > f_c$ is characterized by static electric charges and oscillating director field (dielectric regime). Usually, the two mechanisms of director-electric coupling, rooted in dielectric and conductivity anisotropies, are sufficient to describe electrohydrodynamic instabilities in nematics. In the simplest Carr-Helfrich model of a (-,+) nematic, one considers a balance of a dielectric realigning torque that stabilizes the initial planar orientation in a material with $\Delta\varepsilon < 0$, and an "anomalous" realigning torque caused by a positive conductivity anisotropy $\Delta\sigma > 0$. In the case of directrons, observed in (-,-) nematics with $\Delta\varepsilon < 0$ and $\Delta\sigma < 0$, this classic mechanism is irrelevant, as dielectric and conductivity torques could only stabilize the planar state. The main mechanism of soliton formation is related by Li et al (11, 12) to flexoelectric polarization on the ground of the experimental demonstration that the director oscillates with the frequency of the AC field; the linear model of Earls and Calderer supports this suggestion (62).

The flexoelectric polarization, introduced by Meyer (66),

$$\mathbf{P}_{fl} = e_1\hat{\mathbf{n}}\mathrm{div}\hat{\mathbf{n}} - e_3\hat{\mathbf{n}}\times\mathrm{curl}\hat{\mathbf{n}}, \tag{1}$$

where $e_1$ and $e_3$ are the flexoelectric coefficients, occurs whenever the director suffers splay and bend distortions, which is the case of the directrons, Figs.2-4. Flexopolarization leads to spatial charge density $\rho_{fl} = -\mathrm{div}\mathbf{P}_{fl}$ and a corresponding Coulomb force $\rho_{fl}\mathbf{E}$ (67). Flexopolarization was proposed as a prime mechanism of the so-called "nonstandard" stripe electroconvection in (-,-) nematics (68).

Experimentally observed periodic oscillation of the director in bullets-solitons $B^h$ with the frequency $f$ of the applied electric field (11, 12), Fig.2b, offers direct evidence that the flexoelectric torque $\mathbf{\Gamma}_{fl} = \mathbf{P}_{fl}\times\mathbf{E}$ is responsible for solitons' formation since this torque is linear in the field. A fluctuating director distortion creates a flexoelectric polarization that is enhanced and stabilized by the electric field. The flexoelectric force driving a $B^h$ directron can be estimated roughly as $e^*U \sim 5\times10^{-10}\,\mathrm{N}$, where $e^* \sim 10^{-11}\,\mathrm{C\,m^{-1}}$, is an effective flexocoefficient (5) and $U = 50\,\mathrm{V}$ is the typical voltage (11, 12). This estimate agrees with the expected viscous drag force, $\sim R\eta v \sim 5\times10^{-10}\,\mathrm{N}$, where $R \sim 10\,\mathrm{\mu m}$ is the effective radius of the soliton, $\eta \approx 60\,\mathrm{mPa\cdot s}$ is the viscosity of CCN-47 and $v \approx 0.8\times10^{-3}\,\mathrm{m\,s^{-1}}$ is the directron velocity at $U = 50\,\mathrm{V}$. The speed value corresponds to the high-frequency $B^h$ directrons, in which the flexoelectric polarization is unidirectional and the in-plane director distortion is practically time-independent while the polar angle oscillates with the driving field (11). In the low-frequency $B^l$ directrons, the situation is more complex since the flexoelectric polarization of two asymmetric parts of the solitons are opposite to each other, Fig.4b,c, which leads to somewhat slower propagation.

Recent studies by Aya and Araoka (61) and by Shen and Dierking (13) show that the directrons, apparently of a $B^l$ type, can also propagate in the nematics of the (-,+) type. Aya and Araoka used mixtures of a (-,-) nematic CCN-47 with a (+,+) nematic pentylcyanobiphenyl (5CB). The soliton in these studies are observed only in the mixtures that produce a (-,+) combination, typically driven by the fields of low frequencies 12-20 Hz and of amplitudes less than 10 V; change in the field amplitude redirects solitons, similarly to the case described by Li et al. (12). The director fields of the solitons observed by Aya and Araoka (61) are similar to those established by Li et al for $B^l$ solitons (12), see Fig.3 and Fig.4b,c.

Shen and Dierking (13) showed that the directrons can form not only in the nematics but also in the cholesteric phase. These structures are apparently of both $B^l$ and $B^h$ types since they are observed in a relatively broad range of frequencies, from 10 Hz to 800 Hz. Interestingly, the solitons in the cholesteric phase show some very distinct behavior in collisions, behaving as hardbody particles that cannot pass through each other, apparently because of the mismatch in the overall twisted director (13). Unlike the $B^l$ directrons observed in the nematic phase that have a butterfly structure, Fig.3b,c,e,f, the solitons in the cholesteric phase show a bullet-like appearance.

The unifying feature of directrons observed by various groups is that the conductivity and its anisotropy of the liquid crystal materials are both relatively low. Li et al (11) reported that the conductivity parameters for CCN-47 samples producing $B^h$ solitons were $\sigma \simeq (0.5 - 0.6) \times 10^{-8} \, \Omega^{-1} m^{-1}$ and $\Delta\sigma = \sigma_{\parallel} - \sigma_{\perp} \simeq (-0.1) \times 10^{-8} \, \Omega^{-1} m^{-1}$. To observe $B^l$ directrons, CCN-47 was doped with an ionic additive, to raise the conductivity to $\sigma \simeq (1.2 - 2.5) \times 10^{-8} \, \Omega^{-1} m^{-1}$, Ref. (12). Aya and Araoka (61) observed $B^l$ in the range $\sigma \simeq (0.8 - 4) \times 10^{-8} \, \Omega^{-1} m^{-1}$. Similar conductivity was reported by Shen and Dierking (13) who measured $\Delta\sigma \simeq 1.3 \times 10^{-8} \, \Omega^{-1} m^{-1}$ in the (-,+) material ZLI-2806. Note that all these conductivities and anisotropies are much lower than those typically explored in electrohydrodynamic instabilities in liquid crystals (10).

Nucleation of directrons can be promoted by small particles dispersed in the liquid crystal. Li et al (14) demonstrated that directron can appear at colloidal spheres with tangential boundary conditions. Without the directrons, the spheres have zero electrophoretic activity in an AC electric field. However, when the directrons dress the spheres and thus break the symmetry of the director field around the spheres, the colloids become mobile. The effect is called a directron-induced liquid crystal enabled electrophoresis (14).

To conclude, nematics driven by an AC electric field demonstrate topologically unprotected multidimensional (3+1)D (11) and (3+2)D (12) solitons-directrons of a dissipative type that exist in a certain range of the amplitude and frequency of the field. The time-resolved data on the oscillating director field by Li et al (11, 12) suggest that the directrons are formed because of the flexoelectric polarization coupled to the driving field. The amplitude and frequency of the field control the director structure of directrons, their symmetry, direction of propagation, speed, mutual transformations, disappearance and coalescence into periodic and quasiperiodic patterns. All these features await for an at-depth theoretical description, with the linear model (62) and consideration of charge injection (63) making steps forward.

## II.  LIQUID CRYSTAL-ENABLED ELECTRO-KINETICS

Transport of particles or fluids by an electric field is a subject of electrokinetics. If the driving force represents a uniform electric field, one classifies two closely related sub-areas: electroosmosis, which refers to the dynamics of fluids in contact with solids, and electrophoresis, which refers to the dynamics of dispersed particles in a fluid. Both areas are very active in terms of fundamental science and in practical developments of microfluidics (69, 70), optofluidics (71), small-scale molecular synthesis (72), sensing (73), sorting (74), and biomedical devices (75, 76). Research and applications focus mainly on isotropic electrolytes such as water (77-86) or polymer solutions (87).

Any form of electrokinetics requires separation of electric charges. The effect is well understood when the electrolyte is isotropic. In isotropic electrolytes, the charges are separated by dissociation at the fluid-solid interface or by the different affinity of dissolved positive and negative ions to the solid substrates. If the solid-fluid interface develops an electric double layer in the absence of any electric field, spatial separation of opposite charges in this double layer leads to electro-osmotic flow and electrophoretic mobility in the presence of an electric field. This is the case of linear electrokinetics, in which the flow/particle velocities are growing linearly with the applied electric field. A certain deficiency of linear electrokinetics is that it can be driven only by a direct current (DC) electric field. There is thus an interest in electrokinetic modes in which the velocities grow as the square of the field, as in this case one can use an AC electric field to drive steady transport and flows. In the isotropic electrolyte, these modes, called collectively induced-charge electrokinetics (ICEK), rely on field-induced charge separation. For the field-induced charge separation to happen, the particle or the wall of a fluid cell should have some special properties, e.g., to be an ionic exchanger (80, 81) or a metal (82-86, 88).

Liquid crystals used as anisotropic electrolytes offer a distinct mode of charge separation and of electroosmosis that is called liquid crystal-enabled electrokinetics, or LCEK (89-91). In LCEK, the mechanism of space charge formation is rooted in anisotropy of the medium itself, namely, in anisotropy of electric conductivity and dielectric permittivity, as well as spatially varying molecular orientation director (92). The properties of the transported particle or the nature of the bounding walls are of little importance. Comparison of ICEK and LCEK can be found in Ref.(93).

The essence of LCEK can be understood by considering electro-osmotic flows around a disk-like particle immobilized in a 2D LC cell with a uniform director $\hat{\mathbf{n}}_0$, following Ref. (92). The lateral surface of the particle provides a certain anchoring direction, either normal to itself, Fig.6a, or tangential, Fig.6d. To match the near- and far-fields, the director acquires distortions of a quadrupolar type. Suppose that the dielectric anisotropy is zero, $\Delta\varepsilon = 0$, so that the electric field does not realign the director; its only action is to drive the ions. If $\Delta\sigma > 0$, the ions prefer to move along the director lines. As a result, positive and negative charges gather in different regions of the space. For the disk with normal boundary conditions, the director lines are converging on the left and right sides, Fig.6a. In this case, for $E_x > 0$, the accumulated charges are positive at the left side of the disk and negative on its right side. For the disk with tangential anchoring, the director lines are diverging; the signs of the separated charges in Fig.6d are opposite to that ones in Fig.6a. Comparison of Fig.6a and Fig.6d illustrates clearly that polarity of the space charges is determined by the sign of director gradients. In LCEK, the field-driven charge separation occurs in a deterministic manner, following the pre-existing equilibrium pattern of spatially-varying director, such as the patterns shown in Fig.6. Once the charges are separated in space, the electric field leads to a Coulomb force of density $\propto \rho(E)E$, which yields an electro-osmotic flow of the nematic, Fig.6; here $\rho(E) \propto E$ is the charge density that is proportional to the field that induced it. Reversing the field polarity alters the sign of the induced charge $\rho \propto E$ at a given location, but the product $\rho E \propto E^2$ remains polarity-insensitive. For example, if the field direction shown in Fig.6a is reversed, $E_x < 0$, then the left side of the disk would accumulate negative ions, and the right side would accumulate positive ions; the product $\rho E \propto E^2$ would not change. Therefore the forces and flows are polarity independent, growing as $E^2$. Of course, polarity reversal of charge clouds takes some time, thus the forces and flows decay as the frequency of the AC field becomes higher than the inverse relaxation time.

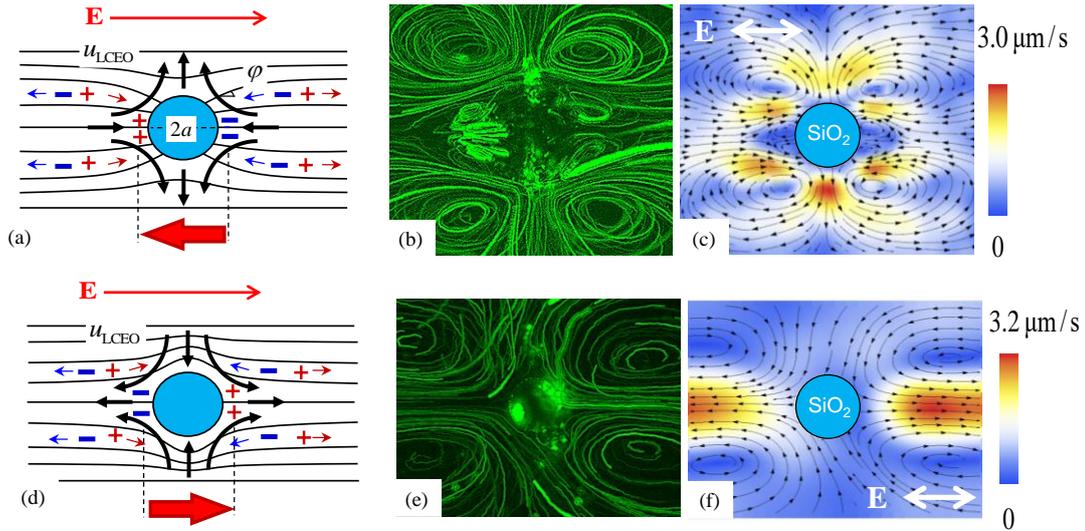

**Fig.6**. Scheme of LCEO in a LC with a director field distorted by a disk with (a,b,c) normal and (d,e,f) tangential anchoring at its surface. In presence of the electric field, anisotropy of conductivity induces space charge. Space separated charges form electric double layers of large spatial extension (comparable to the disk diameter $2a$). The induced dipole moment, shown by red block arrows, is either (a) antiparallel to the applied field or (d) parallel to it, depending on the director gradients. The dipoles reverse their polarity when the electric field is reversed. Electro-osmotic flows of steady directionality are shown by thick arrows in (a,d); their direction does not change when the field polarity is reversed. Corresponding puller- and pusher type experimental LCEO streamlines and velocity maps around glass spheres of diameter 50 $\mu$m with (b),(c) perpendicular anchoring; (e),(f) tangential anchoring (AC field driving of frequency 5 Hz and amplitude 26 $mV/\mu m$). Modified from Ref (91).

The induced space charge is sensitive to the sign of director distortions $\partial \varphi / \partial y$, where $\varphi$ is the angle between the unperturbed $\hat{\mathbf{n}}_0$ and the actual director, Fig.6a. As a result, the two disks with different surface anchoring, Fig.6a and d, produce opposite polarities of flows. The LCEO flow is of a "puller" type around the normally anchored disk with the converging director, i.e., the inward velocities are collinear with $\mathbf{E}$, Fig.6a. The tangentially anchored disk produces a "pusher" pattern of flow, as the inward velocities are normal to the field, Fig.6d.

Experimental verification of the considerations above is illustrated in Fig.6b,c for a sphere with normal anchoring and in Fig.6e,f for a sphere with tangential anchoring. The sphere is immobilized between two glass plates. Application of the electric field results in four main vortices in each quadrant of the system. Since the patterns are left-right and top-bottom symmetric, there is no net flow in the system. If the spheres with a similar director pattern around them were free to move, they would show zero electrophoretic mobility. The situation is changed dramatically when the director field around particles is of a dipolar, Fig.7a, rather than quadrupolar symmetry, as in this case the sphere creates asymmetric vortices, Fig.7b, thus pumping the liquid crystal electrolyte, from right to left, Fig.7c. If such a sphere is not glued, it shows electrophoretic mobility in a uniform AC field with a velocity $v \propto E^2$, directed along the background director (89, 90, 94, 95). If the sphere is of the same material as the one in Fig.7c, it will propel with the point defect, the so-called hyperbolic hedgehog, leading the way (89). The core of the hyperbolic hedgehog is shown in Fig.7a as a filled circle.

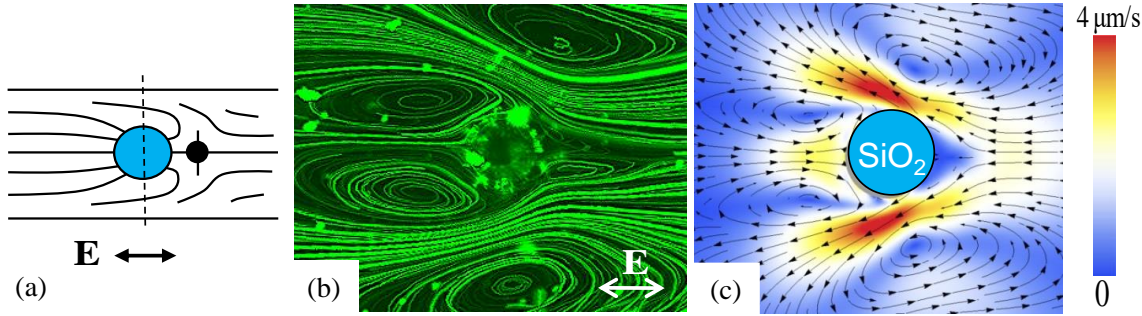

**Fig.7**. Dipolar director field around sphere diameter 50 $\mu m$ with normal boundary conditions glued to a substrate. (a) The point defect, so-called hyperbolic hedgehog, matches the radial director around the sphere and the uniform far field. Broken left-right mirror symmetry causes asymmetric (b) LCEO flows and (c) velocities in a horizontal AC field of frequency 5 Hz and amplitude 26 $mV/\mu m$. The sphere pumps the nematic from right to left. Modified from Ref.(91).

To find how the LCEK velocities depend on the material parameters of the LC, let us consider the simple geometry of Fig.6a,d and estimate the density $\rho$ of charges created as a result of conductivity anisotropy and establish how it depends on the typical scale $a$ (the radius of an inclusion) of director distortions, $E$, and $\Delta\sigma$. Assume that the dielectric anisotropy is small, $\Delta\varepsilon << \bar{\varepsilon}$, where $\bar{\varepsilon} = (\varepsilon_\parallel + \varepsilon_\perp)/2$, and that the 2D director distortions are weak, $\hat{\mathbf{n}} = (1, \varphi)$, where $\varphi = \varphi(x,y)$ is a small tilt angle between the director and the $x$-axis, Fig.6a. The field, applied along the $x$-axis, drives ionic currents $J_i = \sigma_{ij} E_j$, where $\sigma_{ij} = \sigma_\perp \delta_{ij} + \Delta\sigma n_i n_j$ is the conductivity tensor; $i$ and $j$ stand for $x$ and $y$. For $\varphi \ll 1$, the current components are $J_x = \sigma_\parallel E_x + \Delta\sigma \varphi E_y$ and $J_y = \sigma_\perp E_y + \Delta\sigma \varphi E_x$; the field component $E_y$ is induced by separation of charges. Using the charge conservation law $\mathbf{div J} = 0$ and Poisson's equation $\mathbf{div D} = \rho$ (where $\mathbf{D}$ is the electric displacement), one obtains the charge density $\rho(x,y)$ caused by the anisotropy of the LC and director gradients $|\partial\varphi/\partial y| \sim 1/a$ in the presence of an electric field (92):

$$\rho(x, y) = \left(-\frac{\Delta\sigma}{\bar{\sigma}} + \frac{\Delta\varepsilon}{\bar{\varepsilon}}\right)\varepsilon_0\bar{\varepsilon}E_x\frac{\partial\varphi}{\partial y}; \quad (2)$$

here $\bar{\sigma} = (\sigma_{//} + \sigma_{\perp})/2$ is the average conductivity. The induced charge density is thus proportional to the director gradients, the anisotropy of conductivity and permittivity, and the field itself. The amplitude of LCEO velocity around a sphere of radius $a$ then follows from the balance of the driving bulk force $f \propto \rho E$ and viscous resistance $\eta u / a^2$ (92):

$$|u_{LCEK}| = \frac{\beta\varepsilon_0\bar{\varepsilon}a}{\eta}\left|\frac{\Delta\sigma}{\bar{\sigma}} - \frac{\Delta\varepsilon}{\bar{\varepsilon}}\right|E^2. \quad (3)$$

The numerical coefficient $\beta \sim 1$ is introduced to account for the replacement of anisotropic LC viscosity with its average value $\eta$ and for other approximations, such as using $1/a$ as a measure of director gradients.

Theoretical modelling of LCEK and related phenomena has progressed dramatically over the last few years (91, 96-102) but there are still some interesting problems awaiting their analysis. Among these is a seemingly never-completely-resolved problem of the director's dynamic response to the applied electric forcing. In most of the works on LCEK the director field is assumed to be frozen and unresponsive to the electric field. This simplification neglects coupling of the electric field to the director through dielectric anisotropy, flexoelectric and surface polarization. The effect of charge injection from the electrodes is also not accounted for. Although experiments can be designed to supress some of these mechanisms, e.g., by formulating nematic mixtures with $\Delta\varepsilon = 0$ (91, 96), other mechanisms such as flexoelectricity or charge injection are harder to eliminate. Incorporating the director response through dielectric and flexoelectric mechanisms and accounting for charge injection might be the next important steps in expanding mathematical description of LCEK. The interest to further expansion of the LCEK theory is supported by the new experimental results such as omnidirectional propulsion of asymmetric Janus colloidal particles in a nematic driven by the electric field (103) and by potential applications in mixing at microscale (104), sorting and separation (95, 105) of small particles.

## III. LIQUID CRYSTAL ELASTOMERS

A liquid crystal elastomer (LCE) is an anisotropic rubber, as it is formed by cross-linked polymeric chains with rigid rod-like mesogenic segments in the main chain and attached as side branches; these mesogenic units are similar to the molecules forming low-molecular-weight liquid crystals (106, 107). Cross-linked polymeric chains are structurally anisotropic because of their coupling to the orientational order. The coupling enables the mechanical response of LCEs to external factors such as temperature. For example, upon heating, a uniformly aligned LCE strip contracts along the director and expands in the perpendicular directions, since the orientational order weakens and the cross-linked polymer network becomes more isotropic (106). Such a uniform LCE strip behaves as an artificial muscle (108).

Recent research unravels even more exciting effects when the director changes in space, $\hat{\mathbf{n}}(\mathbf{r}) \neq \text{const}$. Thin LCE films with in-plane director patterns develop 3D shape changes with non-trivial mean and Gaussian curvatures when exposed to thermal or light activation (107, 109-117), while director deformations across the film trigger wave-like shape changes and locomotion when activated by light illumination (118). In a parallel vein, there is tremendous progress in exploiting LCE coatings in which one surface is attached to the substrate and the other is free (119, 120). When illuminated with light, photoresponsive coatings with a misaligned director develop random spike-like topographies (121), while periodic elevations and grooves can be produced by using cholesteric "fingerprint" textures (122), or periodic stripe arrays (123). The challenge is in finding an approach by which the change of the topography of the coating or its stretching/contraction can be deterministically pre-programmed.

Modes, Bhattacharya and Warner considered a thin LCE sheet in which the director forms a system of concentric lines (124); the configuration is a topological point defect of strength 1. Upon heating, the LCE shrinks along $\hat{\mathbf{n}}$ by a factor $\lambda < 1$ and expands in the perpendicular directions by a factor $\lambda^{-\nu}$ where $\nu$ is the thermal Poisson ratio. For a circular director field, heating means that the perimeter contracts, $P \to P' = \lambda P$, but the radii extend, $r_0 \to r' = \lambda^{-\nu} r_0$, which could be reconciled only if the initially flat film morphs into a cone. The cone's tip carries localized positive Gaussian curvature $K = 2\pi(1 - \sin\phi)$, where $\phi$ is the cone opening angle (124). Away from the tip, the Gaussian curvature vanishes. However, defects that have topological charge different from 1 could produce Gaussian curvature everywhere around their core (125). These theoretical considerations solve the "forward" design problem, i.e., finding the 3D shape of an activated LCE film when the director field inscribed in its 2D precursor is known. The predicted connection between the topological defects in 2D films and their 3D buckling has been confirmed experimentally, as reviewed by White and Broer (107). A harder task is to solve an inverse design problem, i.e., to find a flat director field that would induce a desired 3D shape upon activation.

A big step forward in generalizing the theoretical insights into design of Gaussian curvature has been made by Aharoni, Sharon and Kupferman (126) who considered smoothly varying director fields. Their paper (126) demonstrated that surfaces of revolution such as spherical, pseudo-spherical and toroidal surfaces, can be produced by patterning the 2D precursor with a smooth director field that depends only on one spatial coordinate. One other interesting feature in Ref. (126) was a recipe to remove degeneracy of 3D "bulging". A given distorted director field could make the 3D shape to bulge either up or down upon heating. If the director is uniform across the film thickness, these two directions are equivalent. However, if the precursor is prepared with the director patterns at the top and the bottom surface slightly twisted with respect to each other, this degeneracy could be removed. The most important message of the work (126) is that the inverse design problem might be solvable even for more general geometries. Aharoni et al (127) combined a numerical approach based on Ref. (126) with advanced experimental control of elastomer preparation, to demonstrate how the inverse design could produce an arbitrary desired shape, such as a face. The desired surface is first presented as a 3D triangulated mesh that is transformed numerically into a 2D mesh with triangles that carry a certain director orientation. The director varies from one triangle to another, as needed for the faithful reconstruction of the 3D shape. At room temperature, the elastomer film is flat. Upon heating, the varying director field causes the triangular mesh to morph into a 3D structure that approximates the desired surface (127).

In a parallel vein, programmed control by a 2D director field was demonstrated to shape dynamic profiles of LCE coatings (128). The program is written as a pattern of molecular orientation in the plane of the LCE coating that is initially flat and rigidly attached to a substrate. When activated by temperature, the coating changes its surface profile as prescribed by the in-plane pattern, by moving the material within and out-of-plane. The displacements are deterministically related to the molecular orientation pattern. For example, upon heating, when the scalar order parameter of the LCE decreases, a circular bend of molecular orientation causes elevations, while a radial splay causes depressions of the coating (128, 129). To understand the relationship, consider how the orientational order is coupled to rubber elasticity of an LCE, as mediated by cross-linking of the polymer network. The coupling results in an anisotropic structure of the network characterized by the so-called step length tensor (106), $l_{ij} = l_{\perp}\delta_{ij} + (l_{\parallel} - l_{\perp})n_i n_j$. The step length $l$ characterizing the polymer segments connecting cross-linking points is different when measured along $\hat{\mathbf{n}}$ ($l_\parallel$) and perpendicularly ($l_\perp$) to $\hat{\mathbf{n}}$. For $l_\parallel > l_\perp$, the spatial distribution of the step lengths can be represented by a prolate ellipsoid elongated along $\hat{\mathbf{n}}$. If the temperature is raised and the orientational order weakens, the distribution becomes more spherical, i.e., the ellipsoid shrinks along $\hat{\mathbf{n}}$ and expands in two perpendicular directions. Once the nematic order is melted, the ellipsoid becomes a sphere, $l_\parallel^{iso} = l_\perp^{iso} = \bar{l}$.

The morphing of the step length ellipsoid caused by the temperature can be modeled by a force dipole, with two point forces of equal amplitude $F$ directed from the poles of the ellipsoid towards its center, Fig.8a. Whenever the director field of the LCE changes in space, so do the local axes of the ellipsoids, Fig.8c,d. The spatial gradients of the step-length tensor produce a vector quantity with the components $f_i = \mu \partial_j n_i n_j$, which can also be written in the equivalent invariant form as (128)

$$\mathbf{f} = \mu\left(\hat{\mathbf{n}}\operatorname{div}\hat{\mathbf{n}} - \hat{\mathbf{n}}\times\operatorname{curl}\hat{\mathbf{n}}\right), \tag{4}$$

where $\mu \propto (l_\perp - l_\parallel)/\bar{l}$ is introduced as an activation parameter that describes the local elastic response to the changing temperature. When the temperature of an LCE with $l_\parallel > l_\perp$ increases and the long axes of the polymer ellipsoids shrink, then $\mu > 0$; in the case of cooling, $\mu < 0$.

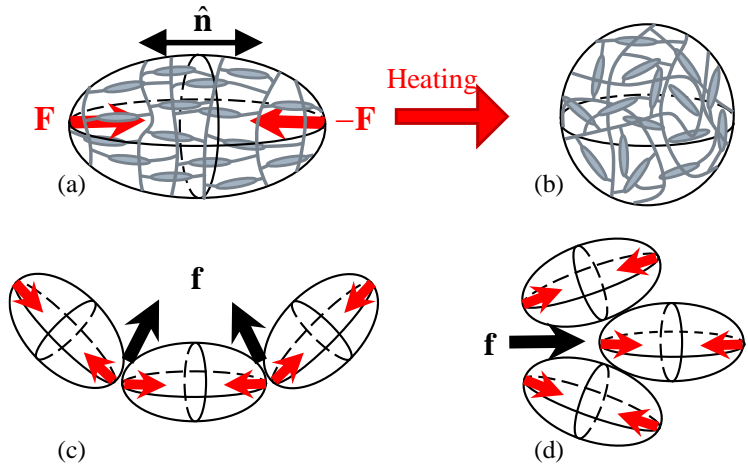

**Fig.8.** Polymer network conformation and occurrence of the activation force. (a) Prolate ellipsoid of polymer network conformations in the nematic phase; the long axis is along the director $\hat{\mathbf{n}}$; during heating, the ellipsoid shrinks along the long axis; in the isotropic phase, it becomes a sphere, as shown in part **(b)**; the shrinking ellipsoid is modelled by a pair of forces $\mathbf{F}$. (c) Activation force density $\mathbf{f}$ produced by contracting ellipsoids in the geometry of bend and (d) splay.

With $\mu$ defined as above, the vector $\mathbf{f}$ represents a spatially varying activation force density that controls the elastic response of an LCE with a non-uniform director $\hat{\mathbf{n}}(\mathbf{r}) \neq \operatorname{const}$ to the external factors such as heating. The occurrence of the force $\mathbf{f}$ is illustrated in Fig.8 for the cases of pure splay and pure bend. For example, in the case of bend, the point forces of the two neighboring shrinking ellipsoids that are titled with respect to each other, produce a net force density $\mathbf{f}$ along the radius of curvature of the director $\hat{\mathbf{n}}(\mathbf{r})$. Coatings with a spatially varying director can also be produced from the smectic precursor; intrinsic undulation instability leads to periodic modulations of the profile that are sufficiently strong to impact the alignment of biological tissues grown on them (130). A useful modification is the development of LCE coatings that swell in the presence of water, forming a nanoscale profile that guides the growth of tissues (131).

The consideration above that led to Eq. (4) illustrates an interesting connection between the out-of-equilibrium behavior of LCE and phenomena such as flexoelectricity, Eq.(1), and Simha-Ramaswamy active force in active matter (132), alluded to in the next section.

**IV MICROSWIMMERS IN NEMATIC**

Motility of microorganisms such as bacteria and spermatozoa is a triumph of evolution. Their locomotion relies on peculiar strategies to overcome viscous drag at low Reynolds number (133) when the inertia forces are negligible, and inspires the development of biological and artificial microbots. Prior research mostly focused on isotropic environments, such as water and water-based solutions (134-136). In an isotropic fluid, the dynamics are typically chaotic, making it difficult to extract useful work, to force the microorganisms to follow prescribed trajectories, to concentrate or disperse them as would be required in technologies of the future such as microscale fabrication, manipulation, delivery, and controllable quorum sensing. In some cases, a sense of direction is set by gradients of nutrients. Unfortunately, these gradients are transient in nature and are therefore not reliable for directing long term transport and useful work. Other channels of communication that served humans so well in domestication of large animals, such as visual or audio cues, are hardly practical when dealing with microorganisms. Similar problems with commanding dynamics arise for artificial microparticles, such as Janus spheres and rods: their propulsion in an isotropic fluid, being ballistic at short time scales, becomes chaotic Brownian-like once the orientational diffusion erases the memory of direction (3, 135, 137).

One of the useful approaches to control microscale dynamics of swimming microorganisms such as bacteria is to use liquid crystals as a structured anisotropic fluid environment. The liquid crystals in question are non-toxic materials based on water dispersions of organic molecules, so-called lyotropic chromonic liquid crystals, or LCLCs (138). The director, either uniform (139-143), or spatially distorted (142-146), serves as an easy swimming pathway for bacteria. Dispersions of bacteria such as *Bacillus subtilis*, in nematic LCLCs, also called "living liquid crystals" or "living nematics", allow one to control independently the activity, through concentration and speed of bacteria (143, 147) and the orientational order, through predesigned director patterns (142),(144, 148), Fig.9. This tunability is important for a better understanding of active systems, as the coupling of the orientational order and activity is at the core of current theoretical modeling (2, 4, 132, 147, 149-156).

*B. subtilis* is a rod-shaped microorganism that could swim in a nematic LCLC at $\sim 15\,\mu m/s$ by coordinated rotation of about 10-20 helical filaments, called flagella; the head of the bacterium rotates in the opposite direction with a smaller frequency (143, 157). *B. subtilis* swimming in a nematic LCLC generates a "hydrodynamic force dipole" pattern (143, 157), typical for many flagellated bacteria swimming in water (158): the surrounding fluid is pushed away from the bacterium along the axial direction and pulled towards the bacterium along the two perpendicular directions. In dense isotropic aqueous dispersions, with a concentration in excess of $10^{15}$ m$^{-3}$, these hydrodynamic dipolar perturbations lead to spectacular collective effects of "bacterial turbulence", with vortices and jets (159, 160). Theoretical modeling of these dense dispersions (161, 162) reveals that the overlapping hydrodynamic force dipoles destabilize the mutual alignment of neighboring bacteria, making a parallel swimming motion unstable. The anisotropic environment of a nematic LCLC can help to rectify the bacterial dynamics and create collective modes of motion (141-144, 148, 157, 163-166), especially when the nematic director is predesigned as a spatially-varying pattern (142, 144, 148, 166).

Of particular interest is to explore predesigned patterns with topological defects that have profound influences on the dynamics, since it is well known that active systems are intrinsically unstable with respect to orientational instabilities that include topological defects (1, 2, 143, 167). A qualitatively simple example is shown in Fig.9, in which the surface-patterned director field (168) forms a +1 topological point defect (144). The director around +1 defects is patterned as a spiral that extends from the center towards the periphery by curling in a clockwise fashion, forming an angle of $\varphi_0 = 45°$ with the radial directions, $\hat{\mathbf{n}} = \{n_r, n_\varphi, 0\} = \{1/\sqrt{2}, -1/\sqrt{2}, 0\}$, where $r$ and $\varphi$ are the radial and azimuthal cylindrical coordinates, respectively. In the language of nematic elasticity, the deformations around the +1 defects are of a mixed splay-bend type when $0 < \varphi_0 < 90°$, pure splay when $\varphi_0 = 0$ and pure bend when $\varphi_0 = 90°$. *B. subtilis* microswimmers show a collective behavior when dispersed in such a cell: once their concentration exceeds some critical value (166), they gather into a swarm around the core of the +1 spiral defect, circulating in a counterclockwise direction, Fig.9b,c.(144) The transition from individual swimming, in which the bacteria simply follow the director field, to their collective unidirectional circulation at 45° to the pre-inscribed director occurs when the number density of bacteria exceeds $c \approx 0.2 \times 10^{13}$ m$^{-3}$, which corresponds to the volume fraction $\Phi \approx 0.6 \times 10^{-5}$ (166). These relatively low concentrations correspond to the onset of the overlapping of hydrodynamic force dipoles of neighboring bacteria, at separation distances (15-20) μm (157). Qualitatively, the collective unidirectional circulation above the concentration threshold can be explained by the active force generated by hydrodynamic interactions of bacteria (144).

Each microswimmer is characterized by a pair of point forces of equal absolute value but opposite polarity (132, 154), similarly to the elastomer scheme in Fig.8a. The collective effect is the active force in the form (132, 154)

$$\mathbf{f}_a = \alpha \left[ \hat{\mathbf{n}} \operatorname{div} \hat{\mathbf{n}} - \hat{\mathbf{n}} \times \operatorname{curl} \hat{\mathbf{n}} \right], \tag{5}$$

where $\alpha < 0$ is the activity (the negative sign reflects the extensile character of *B. subtilis* swimmers, which are "pushers"). For the predesigned director field above, with the spiral angle $\varphi_0 = 45°$, Eq. (3) yields $\mathbf{f}_a = \{0, -\alpha/r, 0\}$ written in the cylindrical coordinates, i.e. the only nonzero component of the active force is the azimuthal component directed counterclockwise, Fig.9d. If the director pattern were purely radial or circular, there would be no unipolar circulation and no net flow in the system (144, 154).

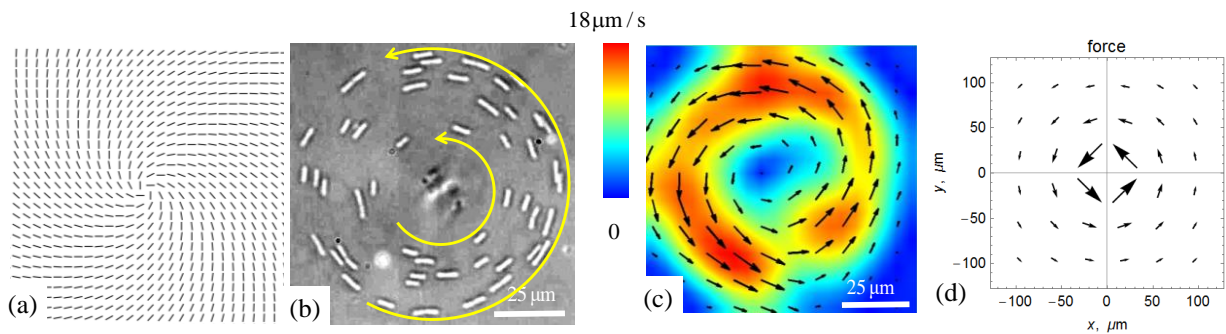

**Fig.9**. (a) Patterned director field with +1 spiral vortex causes (b) unidirectional counterclockwise swimming of a bacterial swarm condensed into a circular band; (c) bacterial velocity map; (d) active force $\mathbf{f}_a$ map.

Koizumi et al (166) extended the study of the collective motion of bacterial swimmers to the patterned vortices of an arbitrary spiral angle, $\hat{\mathbf{n}} = \{\cos\varphi_0, -\sin\varphi_0, 0\}$, in which case the active force $\mathbf{f}_a = \alpha \{\cos 2\varphi_0, -\sin 2\varphi_0\}/r$ shows a nonzero radial component for $\varphi_0 \neq 45°$. When splay dominates the deformation, $0 < \varphi_0 < 45°$, the radial component of the force is directed towards the center of the vortex; the condensed swarm of the bacteria in such a vortex would contract towards the center. When $45° < \varphi_0 < 90°$, a predominantly bend deformation of the vortex moves the swarm towards the periphery.

The simple qualitative consideration above does not account for spatial variations of the bacterial concentration. In theoretical modelling, the active matter is usually considered as incompressible, with the active units as the only ingredient, filling the space fully. As a result, the predicted activity-triggered macroscopic flows involve the entire sample with all its material points. However, in real experimental systems, such as in Fig.9, there is often some space left for the passive background. An intriguing example (169) is active microtubules dispersed in a passive aqueous buffer in a volume proportion 1:1000. As shown by Wu et al (169), when such a dilute, globally isotropic active fluid is confined into a macroscopic 3D channel, it flows in absence of any external pressure gradients, thanks to the formation of a thin active nematic wetting layer at the bounding surface; the bulk remains isotropic. The concentration of active microtubules in this system is thus not necessarily uniform in space. Accounting for the variation in concentration of active units in active matter is one of the current challenges for mathematical modelling. A related problem is the role of splay and bend in active behavior.

Splay and bend affect differently many aspects of active nematics. For example, a system of "pushers", such as the swimming bacteria, at weak levels of activity develops a spontaneous bend (1, 132, 143), while "pullers" are expected to develop a splay (1). Simulations (150, 152, 153) predict that a medium-level activity triggers more complex distortions, such as "kink walls" (152) in which the director field resembles rows and columns of letters "C". In these kink walls, splay (at the ends of C's) and bend (in the middle of C's) coexist and alternate with each other in a plane of the sample. It was also demonstrated (170, 171) that the effective splay and bend elastic moduli are renormalized by activity. In the case of vortices discussed above, the prevalence of splay condenses the swirls of microswimmers towards the core, while the prevalence of bend expands them to the periphery (166).

A different effect of "linear condensation" of active units by splay-bend pattern was presented by Turiv et al (148). In this study, flagellated swimming bacteria are dispersed in a passive nematic LCLC with periodically alternating stripes of splay and bend (148). The passive director is patterned in the form of rows of letters "C" (similar to a stack of parallel Néel wall), $\hat{\mathbf{n}}(x, y, z) = [\cos(\pi y / L), -\sin(\pi y / L), 0]$, where $L$ is the period. Pure splay deformations are located at $y = 0, \pm L, \pm 2L,...$, while bend at $y = \pm L/2, \pm 3L/2,...$. In one assumes $\alpha = \text{const}$, then Eq, (5) leads to $\mathbf{f}_a = (\alpha \pi / L)[-\cos(2\pi y / L), \sin(2\pi y / L), 0]$, which suggests bipolar flows towards the positive ends of the $x$-axis in the splay regions and in the opposite direction in the bend regions. However, in the experiment, the bacteria show a much higher affinity to splay than to bend (148). The splay regions condense pusher-like microswimmers into polar jets moving unidirectionally towards the positive end of the $x$-axis, as prescribed by the polarity of the vector $\alpha \mathbf{s} = \alpha \hat{\mathbf{n}} \text{div} \hat{\mathbf{n}}$, where $\alpha < 0$. In contrast, bend regions produce no net flow and merely guide the bacteria towards the splay regions along the director lines, while helping to quench an undulatory instability of the jets (148). The polar jets are capable of directional cargo transport (148). The acquired knowledge of liquid crystal-mediated control of swimming bacteria might help to develop tools to tune spatio-temporal behavior of microswimmers, to harness the energy of their locomotion, transform it into useful work, use them as parts of micromachines, control biological functions such as quorum sensing and chemotaxis.

The principles of anisotropic medium control of active matter could be extended to other systems. For example, patterned liquid crystal elastomer substrates have been recently used to design tissues of living cells with predetermined locations of topological defects (131). Balance of elastic, surface, and active forces in these engineered tissues allows one to extract useful material parameters such as the Frank elastic constant characterizing the orientational elasticity of the tissue, on the order of 5 nN (131).

**CONCLUSION**

We considered some of the problems in liquid crystals with a spatially-varying director field that might warrant a closer look of mathematicians. All of them are associated with the dynamic phenomena and most deal with complex 3D director structures that evolve in time and space. Examples of electrically driven solitons, the so-called directrons or director bullets, liquid crystal elastomers, and active matter bring a rather unexpected connection through the vector entity, Eqs.(1,4,5), that occurs in spatially distorted director. Proposed initially as flexoelectric polarization by R.B. Meyer (66) and responsible for the formation of directrons, this vector entity acquires a different physical meaning in different contexts, such as the activation force in thermally- or photo-addressed liquid crystal elastomers and in numerous examples of active matter. Recent advances in experimental characterization of a complex director field in 3D suggest that the field is ripe for mathematical modeling. Of immediate interest would be a description of dynamic 3D director fields that accounts for anisotropic bulk elasticity, finite surface anchoring and various coupling mechanisms to the electric field, such as anisotropy of dielectric permittivity, ionic conductivity, flexoelectricity, and surface polarization. Modeling complex director structures out of equilibrium might open a door for new technologies that would use the ability of patterned liquid crystals to harness the energy of microscale dynamics and transform it into useful work.

When this review has been finalized, Noel Clark and his group at the University of Colorado Boulder reported on the discovery of a ferroelectric nematic fluid $N_F$, a new state of matter (172), expected since 1916 (173). $N_F$ is formed by rod-like molecules with strong permanent dipole moments, ~10 D, aligned in a polar fashion, producing a macroscopic electric polarization, which adopts one of the two permissible orientations along the director $\hat{\mathbf{n}}$, $\mathbf{P}$ and $-\mathbf{P}$. $N_F$ is a new distinct member of a broad group of ferroelectrics, most of which are crystalline and thus difficult to mold into patterned polarization fields needed in some applications (174). The fluid character, combined with the polar order of $N_F$ might lead to a plethora of new effects. $N_F$ adds a new dimension to the phenomena discussed in this review, ranging from the response to electromagnetic fields and flows, interplay of bound and free electric charges, fields- and patterns-controlled hydrodynamics, the interplay of polarity and chirality, to the behavior of topological defects (175). Hybrid materials involving $N_F$, such as elastomer liquid crystals would also be of great interest to explore.

# Additional Information

**Acknowledgments**
The author is thankful to the former and current PhD graduate students G. Babakhanova, H. Baza, V. Borshch, Y.-K. Kim, R. Koizumi, I. Lazo, B.X. Li, Yu. A. Nastishin, C. Peng, T. Turiv, S. Zhou, collaborators I.S. Aranson, D.J. Broer, M. C. Calderer, A. Doostmohammadi, M.M. Genkin, D. Golovaty, A.P.H.J. Schenning, A. Sokolov, S.V. Shiyanovskii, J. Selinger, R. Selinger, K. Thijssen, J. Viñals, N.J. Walkington, Q.-H. Wei, J.M. Yeomans, whose contributions made the presented research possible. I also thank the organizers A. Zarnescu, X. Chen, M. Ravnik, and V. Slastikov for the opportunity to present the material in this review at the Spring school on the mathematical design of materials at Isaac Newton Institute for Mathematical Sciences.

**Funding Statement**

The work is supported by NSF DMREF DMS-1729509, NSF DMR-1905053, and by Office of Science, U.S. Department of Energy, grant DE-SC0019105.

**Competing Interests**

*I have no competing interests.*